# Spin-driven evolution of asteroids' top-shapes at fast and slow spins seen from (101955) Bennu and (162173) Ryugu

[1]Masatoshi Hirabayashi, [1]Ryota Nakano, [2,3]Eri Tatsumi, [4]Kevin J. Walsh, [5]Olivier S. Barnouin, [6]Patrick Michel, [7]Christine M. Hartzell, [8]Daniel T. Britt, [3]Seiji Sugita, [9,10]Sei-ichiro Watanabe, [4]William F. Bottke, [11]Daniel J. Scheeres, [12]Ronald-Louis Ballouz, [3]Yuichiro Cho, [3]Tomokatsu Morota, [12]Ellen S. Howell, and [12]Dante S. Lauretta.

[1]Auburn University, Auburn, AL, USA.
[2]Instituto de Astrofísica de Canarias, Tenerife, Spain
[3]University of Tokyo, Tokyo, Japan
[4]Southwest Research Institute, Boulder, CO, USA
[5]The Johns Hopkins University Applied Physics Laboratory, Laurel, MD, USA
[6]Université Côte d'Azur, Observatoire de la Côte d'Azur, CNRS, Laboratoire Lagrange, Nice, France
[7]University of Maryland, College Park, MD, USA
[8]University of Central Florida, Orlando, FL, USA
[9]Nagoya University, Nagoya 464-8601, Japan
[10]Institute of Space and Astronautical Science (ISAS), Japan Aerospace Exploration Agency (JAXA), Kanagawa, Japan
[11]University of Colorado, Boulder, CO, USA
[12]Lunar and Planetary Laboratory, University of Arizona, Tucson, AZ, USA.

## Abstract

Proximity observations by OSIRIS-REx and Hayabusa2 provided clues on the shape evolution processes of the target asteroids, (101955) Bennu and (162173) Ryugu. Their oblate shapes with equatorial ridges, or the so-called top shapes, may have evolved due to their rotational conditions at present and in the past. Different shape evolution scenarios were previously proposed; Bennu's top shape may have been driven by surface processing, while Ryugu's may have been developed due to large deformation. These two scenarios seem to be inconsistent. Here, we revisit the structural analyses in earlier works and fill a gap to connect these explanations. We also apply a semi-analytical technique for computing the cohesive strength distribution in a uniformly rotating triaxial ellipsoid to characterize the global failure of top-shaped bodies. Assuming that the structure is uniform, our semi-analytical approach describes the spatial variations in failed regions at different spin periods; surface regions are the most sensitive at longer spin periods, while interiors fail structurally at shorter spin periods. This finding suggests that the shape evolution of a top shape may vary due to rotation and internal structure, which can explain the different evolution scenarios of Bennu's and Ryugu's top shapes. We interpret our results as the indications of top shapes' various evolution processes.

**Highlights**
- We reanalyzed results from FEM analyses for asteroids Bennu and Ryugu, the targets of OSIRIS-REx and Hayabusa2, respectively.
- We also applied a semi-analytical approach to quantify how the failure modes of top shapes vary with spin.
- Top shapes may evolve by two deformation modes: surface processing at low spin and internal failure at high spin.

**1. Introduction**
Small asteroids are considered to be leftovers generated by re-accumulation of regolith after catastrophic disruptions of their parent bodies [e.g., Bottke et al., 2005, 2015; Morbidelli et al., 2009; Walsh, 2018; Michel et al., 2020]. Near-Earth asteroids (101955) Bennu, a B-type asteroid, and (162173) Ryugu, a Cb-type asteroid, are the targets of the ongoing asteroid sample return missions OSIRIS-REx [Lauretta et al., 2019] and Hayabusa2 [Watanabe et al., 2019], respectively. The recent proximity observations in these missions have given the detailed physical properties of the target asteroids. The unique features of Bennu and Ryugu include the oblate shapes with equatorial ridges, the so-called top-shapes (Figure 1). Re-accumulation processes after catastrophic disruptions may play a key role in the formation of top shapes and their precursor shapes [Michel et al., 2020]. These shapes may have further evolved into the complete form of top shapes at later stages. The circularities of Bennu's and Ryugu's ridges (while the level of such features may differ) suggest that their fast rotation at present or in the past may have contributed to the current configurations of their top shapes. However, the evolution scenarios of their top shapes have been proposed differently.

Bennu's equivalent radius was reported to be 245 m [Barnouin et al., 2019], which is consistent with that estimated by radar observations, 246 m [Nolan et al., 2013]. The spin period was determined as 4.3 h [Nolan et al., 2019; Hergenrother et al., 2019]. The bulk density is 1.19 g/cm$^3$ [Scheeres et al., 2019], consistent with an earlier prediction by analyses of the Yarkovsky drift [Chesley et al., 2014]. This low bulk density, as well as the observed thick regolith layers, implies that Bennu is likely to be a rubble pile [Barnouin et al., 2019; Lauretta et al., 2019; Scheeres et al., 2019; Walsh et al., 2019]. The variation in the surface slope is distinctive at lower and higher latitudes, the boundary of which can be characterized by the size of the Roche lobe at the current spin period [Scheeres et al., 2019]. This feature implies that the surface at lower latitudes is within the Roche lobe and may have evolved due to materials from higher latitudes at present [Scheeres et al., 2016, 2019]. Preliminary observations of the surface morphology found mass movements at multiple scales [Walsh et al., 2019], which may have led to the present boulder distributions [Lauretta et al., 2019] and the surface morphology like small and large grooves [Barnouin et al., 2019]. Such surface diversities imply internal rigidity at low level in the body [Barnouin et al., 2019]. Also, a plastic finite element model (FEM) approach showed that Bennu may not have internal failure but may experience surface failure at the present spin period if the cohesive strength is less than ~0.5 Pa [Scheeres et al., 2019].

Ryugu is a top-shaped asteroid with an equivalent radius of 500 m and is currently rotating at a spin period of 7.6 h [Watanabe et al., 2019]. Ryugu's bulk density is similar to Bennu's, i.e., 1.19 g/cm$^3$ [Watanabe et al., 2019], which implies that this asteroid is also a rubble pile. Observed

OH-bearing materials (but at low level) [Kitazato et al., 2019] and highly porous boulders with low thermal inertia [Okada et al., 2020] imply that Ryugu may be a remnant after catastrophic disruption of an undifferentiated parent body. Also, the Small Carry-on Impact experiment showed that the artificial crater formation is in the gravity regime, implying that the surface is covered by mechanically weak layers [Arakawa et al., 2020]. Unlike Bennu, however, Ryugu's slope does not reflect the current rotational condition. The slope variation becomes the smallest when the rotation period is 3.5 h – 3.75 h. This finding led to an interpretation that there may have been reshaping processes at this spin period. It was found that the surface morphology is less cratered in the western region than that in the eastern region (the east-west dichotomy) [Sugita et al., 2019]. The boulder distribution was also observed to be heterogeneous [Michikami et al., 2019], while mass movements may be limited and correlate with the gravitational potential [Sugita et al., 2019]. Based on these observations, assuming that Ryugu's structure is uniform, Hirabayashi et al. [2019] hypothesized that large deformation occurred at a spin period of 3.5 – 3.75 h to produce the east-west dichotomy.

These studies give some discrepancy of Bennu's and Ryugu's reshaping processes while they have top shapes with a similar bulk density. Specifically, these two bodies are considered to have been affected by rotationally driven deformation but may have evolved differently. Bennu's top shape may have evolved due to surface processing at present. The equatorial ridge may be developed due to mass wasting on the surface while the interior is intact (no mass flows occur in the interior). On the other hand, Ryugu's shape may result from large deformation, perhaps caused by internal failure. The interior becomes the most sensitive to structural failure at high spin, leading to internal mass flows (inward flows along the spin axis and outward flows in the radial direction) that make the shape oblate and push materials on the equatorial plane outward.

We fill a gap between these hypothesized reshaping processes. We revisit the FEM results from earlier studies [Hirabayashi et al., 2019; Scheeres et al., 2019; Watanabe et al., 2019] and explore the variations in the structural failure mode of these objects at different spin periods. To support our discussion, we newly employ a semi-analytical approach proposed by Nakano et al. [2020], who extended an approach by Hirabayashi [2015]. This semi-analytical approach computes Bennu's and Ryugu's yield conditions by assuming their shapes to be uniformly rotating triaxial ellipsoids and characterizes their global failure modes. We finally acknowledge Cheng et al. [2020], who recently proposed that top-shaped asteroids like Ryugu and Bennu result from mass movements in shallow surface layers at a fast rotation period. They assumed a large, rigid core in a target sphere to reproduce a top shape. In the present study, the semi-analytical approach and the FEM approach discuss the variations in the structural failure modes under the uniform structure condition.

We outline the present manuscript. Section 2 briefly reviews the proposed evolution processes of top-shaped asteroids. Section 3 shows the analysis approaches employed in this manuscript: the plastic FEM approach and the semi-analytical approach. Section 4 summarizes the results from our analyses. The results from the FEM approach are already introduced in Hirabayashi et al. [2019], Scheeres et al. [2019], and Watanabe et al. [2019], and we compare these results in this manuscript. Section 5 discusses the evolution of top-shaped asteroids based on the results from Section 4. Our FEM approach assumes the structure of a body to be uniform. On the other hand, the semi-analytical approach provides the distribution of the minimum cohesive strength that can avoid structure failure (see the definition in Section 3) and give insights into the dependence of the failure mode on the internal structure.

## 2. Rotational evolution of top-shaped asteroids

While Section 1 summarized the rotational evolution of top-shaped bodies by focusing on Bennu and Ryugu, this section reviews the general background of such objects. Ground-based radar observations show that spheroidal shapes are dominant in observed objects [e.g., Taylor et al., 2012; Benner et al., 2015]. Among them are reported to be top shapes, some of which also have smaller satellites [e.g., Ostro et al., 2006; Brozović et al., 2011; Becker et al., 2015; Busch et al., 2011; Naidu et al., 2015, 2020; Walsh and Jacobson, 2015]. Some top shapes are rotating close to their spin limits at which surface materials may be about to be shed [Scheeres et al., 2006; Harris et al., 2009], while others like Ryugu are not [Watanabe et al., 2019]. Findings of top shapes in different taxonomy types imply that the evolution of top shapes are not linked to the material compositions of asteroids [e.g., Walsh et al., 2018]. After the formation of top shapes and their precursor bodies during re-accumulation after catastrophic disruption [Michel et al., 2020], these bodies may have evolved by experiencing fast rotation. There are two competing scenarios for rotationally driven reshaping processes that may have induced top shapes.

First, surface processing may be the primary contributor to developing top shapes [e.g., Harris et al., 2009; Minton, 2008; Scheeres, 2015; Walsh et al., 2008, 2012; Cheng et al., 2020]. When the centrifugal force becomes dominant at fast rotation, surface mass movements may occur towards lower potential regions around the equator, while the interior is intact [e.g., Sánchez and Scheeres, 2018, 2020]. The equatorial ridge evolves by the accumulation of surface materials from middle and high latitude regions. When materials slide down to the equatorial region, they may experience the Coriolis force and move towards the longitudinal direction [Harris et al., 2009; Statler et al., 2014]. Also, when surface processing is significant, the contacts of particles may tend to induce more mass ejection on irregular surfaces [Yu et al., 2018].

Second, if the stress field in the interior is considered, another top-shape evolution process may be possible [Hirabayashi and Scheeres, 2015]. With fast rotation, they should have higher stress condition around the center of mass [Hirabayashi, 2015]. This mechanism results from force vectors acting on each element that is affected by gravity and rotation. In a top shape, the direction of the gravity force is radially inward, while that of the centrifugal force is horizontally outward. When the body is rotating at fast spin, the centrifugal force at lower latitudes exceeds the gravity force, leading to a horizontal, outward loading. At higher latitudes, on the other hand, the inward component of the gravitational force is always dominant. These loading modes induce outward deformation in the horizontal direction around the equator and inward deformation in the vertical direction near the spin axis, possibly enhancing the evolution process of a top shape.

A stand-alone explanation of each scenario sounds incomplete to discuss the top-shape evolution mechanism. If surface processing is a main driver that develops top shapes at fast spin, the interior should be mechanically stronger than the surface layers. On the other hand, if the mechanical strength of the interior is similar to that of the surface layer, the stress field in the interior should reach its yield condition before surface regions experience mass movements [Hirabayashi, 2015; Hirabayashi et al., 2015]. Gravity measurements implied that Ryugu's structure may be homogeneous [Watanabe et al., 2019]. The present study attempts to explain the correlation of surface and internal failure by considering the structural conditions at different spin periods.

## 3. Analysis approach
### 3.1. Results from the earlier plastic FEM approach

A plastic FEM approach was developed and designed to analyze the locations of structurally failed regions in an irregularly shaped body that is uniformly rotating. The developed code was compatible with ANSYS Mechanical APDL 18.1. In this study, instead of conducting new FEM analyses, we revisit the structural analysis results for Bennu and Ryugu from our earlier works [Watanabe et al., 2019; Scheeres et al., 2019; Hirabayashi et al., 2019]. We show how Bennu and Ryugu experience rotationally driven failure modes at multiple spin periods when the structure is assumed to be uniform. While the present work does not provide new simulations, we briefly introduce our FEM simulation settings below. Note that the detailed model development is also provided in the earlier work [e.g., Hirabayashi et al., 2016; Hirabayashi and Scheeres, 2019].

The FEM meshes of Bennu and Ryugu were generated from the mission-derived polyhedron shape models [Barnouin et al., 2019; Watanabe et al., 2019]. We first used TetGen [Si, 2015] to convert polyhedron shape models to 4-node triangular FEM meshes. Then, these FEM meshes were modified to obtain 10-node meshes. Because the meshes developed by TetGen were not optimally compatible with the simulation environment in ANSYS, we refined them by performing manual operations and by using the ANSYS mesh refining functions. The derived mesh was further processed to provide the constraints on the spatial movements of nodes. To mimic the condition of an irregularly shaped body in space, we gave constraints on three degree of freedom at the center of gravity for the translational motion and another three degrees of freedom at two surface nodes for the rotational motion [Hirabayashi and Scheeres, 2019]. We computed a force loading acting on each node by accounting for the self-gravity force and the centrifugal force.

Our model assumes that structural deformation is small and follows linear elasticity below the yield condition and perfect plasticity on that condition. Perfect plasticity specifies that there is no material hardening and softening. To define the yield condition, we used the Drucker-Prager yield criterion, which is written as [Chen and Han, 1988]

$$f = \alpha I_1 + \sqrt{J_2} - s \leq 0. \tag{1}$$

$I_1$ and $J_2$ are the stress invariants, which are given as

$$I_1 = \sigma_1 + \sigma_2 + \sigma_3, \tag{2}$$

$$J_2 = \frac{1}{6}\{(\sigma_1 - \sigma_2)^2 + (\sigma_2 - \sigma_3)^2 + (\sigma_3 - \sigma_1)^2\}, \tag{3}$$

where $\sigma_i (i = 1, 2, 3)$ is the principal stress component. $\alpha$ and $s$ are material constants that are defined as

$$\alpha = \frac{2 \sin \phi}{\sqrt{3}(3 - \sin \phi)}, \quad s = \frac{6Y \cos \phi}{\sqrt{3}(3 - \sin \phi)}, \tag{4}$$

where $\phi$ and $Y$ are the angle of friction and the cohesive strength, respectively. We note that $Y$ is defined in the sense of continuum media, which is the cumulative effect of cohesion caused by grain-grain interactions in a given small element.

In our FEM simulations, the structural condition was assumed to be uniform over the entire body. Young's modulus ($E_s$), Poisson's ratio ($\mu$), and the angle of friction ($\phi$) were fixed at $1 \times 10^7$ Pa, 0.25, and 35°, respectively [e.g., Lambe and Whitman, 1969]. For simplicity, no dilatancy was allowed in these simulations. The necessary parameters for FEM simulations are available in Table 1. To determine the structural failure mode of an irregularly shaped body at a given spin period, we changed the cohesive strength under the constant parameters. If the actual cohesive strength is enough to resist failure, the body can keep the original shape. However, as the cohesive strength becomes smaller, some regions in the body start failing structurally. We computed the distribution of failure and the minimum cohesive strength that the body can avoid failure anywhere (later denoted as $Y^*$) [Hirabayashi and Scheeres, 2019].

*Table 1. Physical parameter settings in simulations. The dimensions are from Barnouin et al. [2019] for Bennu and Watanabe et al. [2019] for Ryugu.*

| Physical Parameters | Symbols | Value | Units |
|---|---|---|---|
| Young's modulus | $E_s$ | $1 \times 10^7$ | [Pa] |
| Poisson's ratio | $\nu$ | 0.25 | [-] |
| Angle of friction | $\phi$ | 35 | [deg] |
| Bulk density | $\rho$ | 1.19 | [g/cm3] |
| Bennu's dimension | $[2a, 2b, 2c]$ | $506 \times 492 \times 457$ | [m$^3$] |
| Ryugu's dimension | $[2a, 2b, 2c]$ | $1,004 \times 1,004 \times 876$ | [m$^3$] |

### 3.2. Semi-analytical stress model

While our plastic FEM approach can describe the failure modes of irregularly shaped bodies, the results of structural failure depend on local topographic features [Hirabayashi and Scheeres, 2019]. Such local features may be too detailed to study global conditions that induce either surface processing or internal failure. Here, to characterize the mechanisms of such global failure modes, we apply a semi-analytical approach that computes the variations in the stress distributions in a uniformly rotating triaxial ellipsoid at different spin periods. Note that the model assumption of a rotating triaxial ellipsoid gives a reasonable approximation for the global failure modes of a top-shaped asteroid. The present approach is an extension of the previous approach, which focused on a spherical body [Hirabayashi, 2015]. This approach is first discussed by Nakano and Hirabayashi [2020].

The semi-analytical approach consists of two aspects. The first aspect is to compute the stress distribution in a uniformly triaxial ellipsoid by assuming linear elasticity [Dobrovolskis, 1982; Holsapple, 2001; Love, 2011]. The second aspect is to use this stress distribution and the Drucker-Prager yield criterion to derive the minimum cohesive strength, $Y_e^*$, at given elements. We distinguish $Y_e^*$ with $Y^*$, which is given as the minimum cohesive strength that a considered body starts to experience failure somewhere (from Section 3.1). Importantly, we define the following relationship: $Y^* = \max(Y_e^*)$. This approach does not provide the deformation processes but shows the sensitivity of a body element to structural failure. We confirmed that the earlier semi-analytical approach and the FEM approach gave consistent results for the failure modes of a uniformly rotating spherical body [Hirabayashi, 2015]. Below, we describe how to compute $Y_e^*$. The details are also found in Nakano and Hirabayashi [2020].

For the first aspect, the stress distribution in a uniformly rotating triaxial ellipsoid is computed based on linear elasticity [Dobrovolskis, 1982; Holsapple, 2001; Love, 2011]. Considering the equilibrium state of stress, we can write the stress distribution as

$$\frac{\partial \sigma_{11}}{\partial x_1} + \frac{\partial \sigma_{12}}{\partial x_2} + \frac{\partial \sigma_{13}}{\partial x_3} + \rho b_1 = 0, \qquad (5)$$

$$\frac{\partial \sigma_{21}}{\partial x_1} + \frac{\partial \sigma_{22}}{\partial x_2} + \frac{\partial \sigma_{23}}{\partial x_3} + \rho b_2 = 0, \tag{6}$$

$$\frac{\partial \sigma_{31}}{\partial x_1} + \frac{\partial \sigma_{32}}{\partial x_2} + \frac{\partial \sigma_{33}}{\partial x_3} + \rho b_3 = 0, \tag{7}$$

where $\sigma_{ij}$ ($i,j = 1,2,3$) is the stress component, $\rho$ is the bulk density, which is assumed to be uniform over the entire volume, and $b_i$ ($i = 1,2,3$) is the body force, which includes a centrifugal force and a self-gravity force [Holsapple, 2001]. In these equations, $\sigma_{12} = \sigma_{21}$, $\sigma_{32} = \sigma_{23}$, and $\sigma_{13} = \sigma_{31}$. Because a triaxial ellipsoid is axisymmetric, we define the $x_1$, $x_2$, and $x_3$ axes along the minimum, intermediate, and maximum moment of inertia axes, respectively.

From Equations (5) through (7), the displacement $u_i$ in the $x_i$ direction is expressed with 12 parameters [Dobrovolskis, 1982],

$$u_1 = x_1 \left[ A + B \frac{x_1^2}{a^2} + C \frac{x_2^2}{b^2} + D \frac{x_3^2}{c^2} \right], \tag{8}$$

$$u_2 = x_2 \left[ E + F \frac{x_1^2}{a^2} + G \frac{x_2^2}{b^2} + H \frac{x_3^2}{c^2} \right], \tag{9}$$

$$u_3 = x_3 \left[ I + J \frac{x_1^2}{a^2} + K \frac{x_2^2}{b^2} + L \frac{x_3^2}{c^2} \right], \tag{10}$$

where $a$, $b$, and $c$ are the semi-major axis, the semi-intermediate axis, and the semi-minor axis, respectively ($a \geq b \geq c$). Also, the upper alphabets indicate coefficients that are determined by the constitutive law and the boundary condition. Using Equations (8) through (10), we can describe the strain components as

$$\epsilon_{ij} = \frac{1}{2} \left( \frac{\partial u_i}{\partial x_j} + \frac{\partial u_j}{\partial x_i} \right). \tag{11}$$

We next consider Hooke's law (linear elasticity), which is given as

$$\sigma_{ij} = \lambda \epsilon_{kk} \delta_{ij} + 2\mu \epsilon_{ij}, \tag{12}$$

where $\epsilon_{kk} = \epsilon_{11} + \epsilon_{22} + \epsilon_{33}$ and $\delta_{ij}$ is the Kronecker delta. $\lambda$ and $\mu$ are Lame's constants, which are given as

$$\lambda = \frac{E_s \nu}{(1 + \nu)(1 - 2\nu)}, \tag{13}$$

$$\mu = \frac{E_s}{2(1 + \nu)}. \tag{14}$$

The traction boundary condition is given as

$$\sigma_{ij} n_j = 0, \tag{15}$$

where $n_j$ is written as

$$n_1 = \frac{x_1}{a^2\chi}, n_2 = \frac{x_2}{b^2\chi}, n_3 = \frac{x_3}{c^2\chi}, \qquad (16)$$

and

$$\chi = \left(\frac{x_1^2}{a^4} + \frac{x_2^2}{b^4} + \frac{x_3^2}{c^4}\right)^{\frac{1}{2}}. \qquad (17)$$

Combining all these conditions, we obtain 12 linear equations to uniquely determine the 12 coefficients appearing in Equations (8) through (10).

For the second aspect, we convert these stress components to the principal stress components, $\sigma_1$, $\sigma_2$, and $\sigma_3$. Then, we substitute these principal components into Equation (1). Using the constant parameters defined in Table 1, we compute the distribution of $Y_e^*$ in a uniformly triaxial ellipsoid at different spin periods. Specifically, $Y_e^*$ is computed by considering $f = 0$ in Equation (1).

## 4. Comparison of the structural conditions between Bennu and Ryugu
### 4.1. Results from finite element analysis approach

We first summarize the failure modes of Bennu and Ryugu at different spin periods by introducing the earlier results from our plastic FEM approach [Hirabayashi et al., 2019; Scheeres et al., 2019; Watanabe et al., 2019]. The present study shows the dependence of the failure mode on the spin period, which was also predicted for irregularly shaped bodies by Hirabayashi and Scheeres [2019]. If Bennu and Ryugu are rotating at longer spin periods, their failure mode is characterized as surface failure. On the other hand, if they are rotating at shorter spin periods, internal failure becomes significant. We describe the contrasts of these processes by referring to the plastic FEM results [Hirabayashi et al., 2019; Scheeres et al., 2019; Watanabe et al., 2019].

Figure 2 shows the failure modes of Bennu at spin periods of 4.3 h (the current spin period) and 3.5 h [Scheeres et al., 2019]. At the 4.3 h spin period, the interior does not fail structurally, while the surface regions at middle and high latitudes are sensitive to structural failure (Panels a and b). This failure mode appears when the cohesive strength is ~0.2 Pa. If the cohesive strength of surface materials is below this level, mass movements should occur. The flow processes observed by OSIRIS-REx [Barnouin et al., 2019; Scheeres et al., 2016, 2019; Walsh et al., 2019; Jawin et al., 2020] imply that tin top-layers may be covered with cohesionless materials. Thus, our results support that Bennu's top shape may have evolved by surface processing at the present spin period. At the 3.5 h spin period, on the other hand, Bennu should experience internal failure if the cohesion is less than 0.8 Pa. The central region is the most sensitive to structural failure; the predicted failed region is almost axisymmetric and reaches the top surface.

Ryugu's structural behavior is similar to Bennu's, although the variations in the failed regions appear due to local topography at some level (Figure 3). At a spin period of 7.6 h (the current spin period), the structure does not reach the yield condition in the almost entire volume when the cohesive strength is fixed at 1 Pa (Panels a and b). This trend results from self-gravity, which leads the interior to have compression with limited shear and thus not to experience structural failure significantly. At a spin period of 3.5 h, internal failure becomes significant [Watanabe et al., 2019]. Unlike the symmetric failed region appeared in Bennu at the 3.5 h spin period, Ryugu's failed region is asymmetric (Panels c and d) [Hirabayashi et al., 2019]. Structural failure is observed to be more concentrated in the central area and in the bottom-right part of the plots. This asymmetric distribution of the failed region may imply that in the current shape, Ryugu's internal structure at the 3.5 h spin period is divided into a structurally relaxed region (the upper-left part) and an unrelaxed region (the bottom-right part).

From these structural conditions at different spin periods, the evolution process of Ryugu's top shape was interpreted to be different from that of Bennu's. Ryugu's top shape may have evolved by large deformation when this asteroid was rotating at a spin period of 3.5 ~ 3.75 h [Watanabe et al., 2019]. Because of Ryugu's east-west dichotomy, the surface in the western region is smoother than that in the eastern region [Sugita et al., 2019]. Also, the equatorial ridge in the western region is sharper than that in the eastern region [Watanabe et al., 2019]. The locations of the failed regions predicted from the plastic FEM approach correspond to those of the eastern region [Hirabayashi et al., 2019]. Therefore, the relatively less cratered condition of the western region implies that this region may be structurally relaxed due to a large deformation process at a spin period of 3.5 h ~ 3.75 h [Hirabayashi et al., 2019]. We note that when Ryugu's spin has slowed down to the current spin period, 7.6 h, the body may have also had surface processing as Bennus has.

### 4.2. Results from semi-analytical approach

The FEM approach gave a significant interpretation that Bennu's top shape may result from surface processing at the current spin period, 4.3 h, while Ryugu's may be developed by large deformation, which is due to internal failure, at a spin period of 3.5 h ~ 3.75 h. This contrast between Bennu's and Ryugu's failure modes implies that top shapes may be able to evolve by both surface processing and internal deformation. This section uses the semi-analytical approach, which can describe global failure modes, to further explore how these failure modes appear at different spin periods. We compute the distribution of $Y_e^*$ over slices of Bennu and Ryugu along their spin axis. We incorporate the parameter conditions defined in Table 1 into the semi-analytical model.

First, we discuss the distribution of $Y_e^*$ for Bennu in Figure 4. We observe that the distribution of $Y_e^*$ varies as a function of the spin period. When the spin period is shorter than 3.5 h, $Y_e^*$ becomes the maximum at the center of the body. At a spin period of 3 h, for example, $Y_e^*$ at the central region reaches higher than 12 Pa. $Y_e^*$ on the surface in the equatorial region is lower than that in the central region, but still higher than 0 Pa. In the polar regions, $Y_e^*$ becomes 0 Pa, implying that there is no strength necessary to avoid structural failure. This result indicates that the interior is the most sensitive to structural failure if the actual cohesive strength is uniform over the body. At a spin period longer than 3.75 h, nonzero $Y_e^*$ is concentrated in the equatorial surface region but does not appear in the central region. As the spin period continues to become longer, the regions with nonzero $Y_e^*$ become smaller. At a spin period of 4.3 h (the current spin period), nonzero $Y_e^*$ only appears in shallow surface regions at the equator. In this region, $Y_e^*$ is less than 1 Pa. This result indicates that if the actual cohesive strength is less than that level, surface materials should slide down to regions at lower latitudes. This interpretation is consistent with what was observed from the plastic FEM approach (Figure 2).

Next, we show the distribution of $Y_e^*$ for Ryugu. The bulk density of Ryugu is the same as that of Bennu. Their global shapes are also similar. Therefore, we anticipate that Ryugu's failure mode is similar to Bennu's. Figure 5 illustrates the failure modes of Ryugu at spin periods from 2.5 h to 7.6 h. The magnitude of $Y_e^*$ for Ryugu is higher than that for Bennu because of the size difference. At a spin period shorter than 3.5 h, the region around the center needs higher $Y_e^*$ to resist structural failure. However, when the spin period is around 3.75 h, the region with nonzero $Y_e^*$ is separated towards the equatorial surface regions. As the spin period becomes longer, this region shrinks and eventually disappears. At a spin period of 7.6 h, there is no obvious region with nonzero $Y_e^*$. This result implies that there is no significant deformation at the current spin period. Again, the results are consistent with those from the plastic FEM approach (Figure 3).

Our semi-analytical model describes that if the structure is uniform, structural failure in Ryugu and Bennu can be characterized by two modes. The first mode is that the interior structurally fails first at fast spin. This mode can induce vertical compression along the spin axis and radial deformation in the horizonal direction. The second mode is that surface processing is dominant, and top-surface layers at low and middle latitudes fail structurally at low spin. For these failure modes, the size of internal failure is larger than that of surface failure. As the spin period becomes longer, the transition occurs when nonzero $Y^*$ disappears from the central region and is split towards the surface regions. While these modes are apparently different, they induce mass movements towards the equatorial region and thus can generate top shapes. The similar finding was presented by Hirabayashi [2015] for a limited case in which a body is perfectly spherical.

## 5. Discussion
### 5.1. Surface processing and internal deformation can develop a top-shaped body

The results in Section 4 showed that both surface processes and internal deformation can contribute to the evolution of Bennu's and Ryugu's top shapes. At a longer rotation period, surface layers are the most sensitive to structural failure, while an interior is structurally intact. If surface layers do not have enough cohesive strength, they should be modified. As the spin period becomes shorter, the area of elements sensitive to structural failure increases. Eventually, the central region becomes the most sensitive. At this stage, the interior fails structurally, leading to a deformation mode that consists of a horizontal, outward flow and a vertical, inward flow. The development of a top shape due to these two processes was confirmed by studies that used Discrete Element Model [e.g., Hirabayashi et al., 2015; Sánchez and Scheeres, 2016, 2018, 2020; Walsh et al., 2008, 2012; Yu et al., 2020; Zhang et al., 2017, 2018].

The dependence of deformation on the spin period controls the evolution of a top shape. Because the centrifugal force is the main contributor to reshaping, the shape evolves so that its moment of inertia along the spin axis always increases (Section 4). Because the angular momentum is constant during such a deformation process, the spin period becomes longer. On the other hand, YORP may accelerate/decelerate the spin state, depending on the shape and how these effects have affected the body [e.g., Bottke et al., 2015; Statler, 2009, 2015]. If top shapes spin up, they again become sensitive to another deformation process.

The magnitude of reshaping may be controlled by the evolution of the spin period. In Figure 6, we show the schematics of possible paths of the evolution process of a top shape. When the body is spinning at a spin period at which internal failure occurs (the upper path), deformation is large. This deformation process leads to a higher moment of inertia under constant angular momentum, making the spin period longer. After this event, deformation stops. To have another deformation process, the body needs to be rotationally accelerated for a long time. On the other hand, if the body is spinning at relatively slow spin, surface processes can still occur (the lower path), and the magnitude of deformation may be small. Because the rotational change may be small during such an event, it is not necessary for the body to wait for a long spin-up duration to have another small deformation event. This implies that surface processing events occur more frequently than internal deformation. The paths can interact with each other. This evolution process indicates a complex coupling relationship between the shape and the spin period evolution [Cotto-Figueroa et al., 2015]. Also, we point out that deformation is not necessarily axisymmetric. Structural and shape heterogeneity even at small level can induce asymmetric deformation processes [Hirabayashi et al., 2015; Sánchez, 2015; Sánchez and Scheeres, 2016]. When a structurally weak location experiences deformation, that region should be relaxed structurally. Then, the body that had deformation in one area now has structural sensitivities in other areas, which may deform later.

Bennu has complex trends of mass movements [Walsh et al., 2019; Jawin et al., 2020] and unique morphologies such as grooves [Barnouin et al., 2019]. The energy analysis showed that the boundary of the Roche lobe at the current spin period would be consistent with the surface slope variation observed [Scheeres et al., 2016, 2019]. These findings imply that Bennu's surface topography is currently evolving. The present study implies that at the current spin period, Bennu's interior should be structurally intact, while its surface is sensitive to structural failure (Figure 4). By "intact", we mean that $Y_e^*$ in the interior is zero, thus deformation driven by rotation may not occur. This condition leads to surface processing from the middle latitudes to the equatorial regions if the strength is low. Each mass movement event is small and asymmetric. Because the change at the spin period is small, deformation processes occur occasionally. Also, Bennu's not-perfectly-circular equatorial ridge imply that the spin period has not reached 3.5 h, and the ridge has evolved within the gravity force-dominant region, i.e. the inside of the Roche lobe. If Bennu had experienced a shorter spin period than 3.5 h, the ridge would have been modified to be circular due to a strong centrifugal force [Scheeres et al., 2016, 2019].

Ryugu's top shape, on the other hand, may have evolved due to large deformation. This idea was supported by its almost-perfectly-circular equatorial ridge. Watanabe et al. [2019] pointed out that such a circular ridge might indicate the centrifugal force-gravity force boundary that has occurred at a critical spin period, 3.5 h. At this spin period, the interior of this asteroid becomes the most sensitive to structural failure (Section 4). Hayabusa2 showed that the western region looks apparently younger than the eastern region [Sugita et al., 2019]. This east-west dichotomy implies different time scales for the reshaping process [Hirabayashi et al., 2019]. First, a large-scaled deformation may have occurred to create the global feature of Ryugu's top shape (such as its equatorial ridge). Then, another large-scaled deformation may have happened in the western region and developed its young, smooth terrain.

From these analyses, we conclude that both internal deformation and surface processing can lead to top shapes. After re-accumulation processes due to catastrophic disruptions [Michel et al., 2020], precursor bodies experience these processes at different spin periods in complex ways to evolve into top shapes. Again, it was previously argued that top shapes would be attributed to one of these processes only.

### 5.2. Insights into reshaping and top shapes' interiors from surface morphologies

While the present study assumed that the structure of Ryugu and Bennu are homogeneous, small bodies may have local heterogeneity at some level, given their formation and evolution processes. Many bodies smaller than 100 km in diameter may result from the collisional evolution of larger bodies [Bottke et al., 2005, 2015; Morbidelli et al., 2009; Walsh, 2018]. The following re-accumulation processes may cause the heterogeneities of rubble pile asteroids [Michel et al., 2020]. Even after re-accumulation, deformation processes can change their volumes, changing the density distribution [e.g., Hirabayashi and Scheeres, 2015; Scheeres et al., 2019; Bagatin et al., 2020]. Granular convections may modify the structure of asteroids [e.g., Murdoch et al., 2013; Yamada and Katsuragi, 2014]. Seismic shaking may be a driver that induces granular convection. We note that seismic waves may be significantly attenuated in granular media in a microgravity environment [Yasui et al., 2015; Matsue et al., 2020], as seen by the Hayabusa2 SCI impact experiment [Arakawa et al., 2020, Nishiyama et al., 2020]. Hypervelocity impact processes also resurface and reorganize the interior, depending on their energy levels [e.g., Wünnemann et al., 2006].

The failure mode of a top-shaped asteroid depends on such structural heterogeneities [Hirabayashi, 2015]. Based on the distributions of $Y_e^*$ in Ryugu and Bennu (Figures 4 and 5), Figure 7 describes the dependence of the failure mode on the cohesive strength. This schematic was first introduced by Hirabayashi [2015]; in the present study, we also add the case of a weak interior proposed by Sánchez and Scheeres [2018]. If the structure is uniform, both surface processing and internal failure can contribute to reshaping. When structurally weaker layers are placed on top of a stronger interior, only surface particles can move at high spin [Hirabayashi et al., 2015; Sánchez and Scheeres, 2016, 2018, 2020; Yu et al., 2020]. If the interior is weaker than the surface layer, complex reshaping may occur to generate irregularly shaped bodies rather than top shapes [Sánchez and Scheeres, 2018].

The observed crater morphologies may imply local heterogeneity of Bennu and Ryugu. On Bennu, some small craters exhibit the discontinuities of the boulder size distribution inside and outside the rims [Walsh et al., 2019]. Similar features are also observed on Ryugu [Sugita et al., 2019; Watanabe et al., 2020; Morota et al., 2020]. Small particles are distributed inside such craters, while the outside exhibits larger boulders [Walsh et al., 2019]. One explanation may be grain segregation due to granular convection such as the Brazil nut effect [Watanabe et al., 2020]. Another explanation may be that small particles move on top surface layers more favorably than large boulders and are depleted [Szalay et al., 2018]. If this is the case, top surface layers can be considered to be mechanically weak and have evolved rapidly. Such transportation may be driven not only by combinations of the centrifugal force and the gravitational force but also by other mechanisms. An example may be electrostatic lofting, the amount of which is dependent on cohesion and large electrostatic forces driven by photoemission or secondary electron emission [Hartzell, 2019; Hartzell et al., 2020; Wang et al., 2016; Zimmerman et al., 2016]. Micrometeoroid impacts [e.g., Sugita et al., 2019; Walsh et al., 2019] and thermal cracking [e.g., Molaro et al., 2015, 2017, 2020; Hazeli et al., 2018] may also result in small particle transportation.

### 5.3. Necessary technical improvements

We finally discuss that the current study has two issues that need to be solved in our future work. First, our discussions assumed that the body has a uniform structure. However, as discussed in Section 5.2, there may exist some structural heterogeneity. If this is the case, we expect the variations in the internal structure. Such variations affect our simulation settings, e.g., the friction angle and the density distribution. Second, our model does not give detailed local failure modes, which may highly correlate with the timescale that materials have been exposed to space weathering. In carbonaceous asteroids, space weathering may make materials bluer and redder, depending on their conditions [Gillis-Davis et al., 2017; Hendrix et al., 2019; Lantz et al., 2017; Matsuoka et al., 2015; Thompson et al, 2019]. Recent studies pointed out that on both Bennu and Ryugu, space weathering may redden materials [Sugita et al., 2019; Morota et al., 2020; DellaGuistina et al., 2019, 2020] while space weathering on Bennu needs further interpretations. However, the distribution of bluer materials is highly localized, and our models are too coarse to provide consistent results. These issues need to be resolved to give stronger insights into the evolution mechanisms of top-shaped asteroids.

## 6. Conclusion

The present study explored the rotationally driven evolution of Bennu's and Ryugu's top shapes. We started by revisiting the previously reported results derived from a plastic FEM approach. The semi-analytical approach was then introduced to describe the minimum cohesive strength distribution that can avoid structural failure at a given element, under the assumption that a top shape is a uniformly rotating triaxial ellipsoid. We then quantified how the failure modes of Bennu and Ryugu would vary at different spin periods. The results showed that the failure modes depended on the spin period. For these bodies, surface regions would be sensitive to structural failure at a long spin period, while interiors could be the most vulnerable at a higher spin period. Therefore, we suggest that top shapes can evolve by various reshaping processes that depend on both rotation and internal structure.

## Acknowledgments

M.H. and R.N. thank support from NASA/Solar System Workings (NNH17ZDA001N/80NSSC19K0548) and Auburn University/Intramural Grant Program. Also, the FEM results shown in this manuscript are obtained by using ANSYS Mechanical APDL 18.1. P.M. acknowledges funding support from the French space agency CNES, from the European Union's Horizon 2020 research and innovation programme under grant agreement No 870377 (project NEO-MAPP) and from Academies of Excellence: Complex systems and Space, environment, risk, and resilience, part of the IDEX JEDI of the Université Côte d'Azur. S.S. acknowledges support from Japan Society for the Promotion of Science (JSPS) Core-to-Core Program "International Network of Planetary Sciences".

Watanabe, S. et al. (2019), Hayabusa2 arrives at the carbonaceous asteroid 162173 Ryugu – A spinning top-shaped rubble pile, Science, 364, 6,437, p.268-272, doi:10.1126/science.aav8032.

Watanabe, S. et al. (2020), Paucity of boulders in shallow craters on asteroid 162173 Ryugu, 51th Lunar and Planetary Science Conference, Houston, Texas, 1675.

Wünnemann, K. et al. (2006), A strain-based porosity model for use in hydrocode simulations of impacts and implications for transient crater growth in porous targets, Icarus, 180, p.514 – 527, doi:10.1016/j.icarus.2005.10.013.

Yamada, T. M. and H. Katsuragi, Scaling of convective velocity in a vertically vibrated granular bed, Planetary and Space Science, 100, p.79 – 86, doi:10.1016/j.pss.2014.05.019.

Yasui, M et al. (2015), Experimental study on impact-induced seismic wave propagation through granular materials, Icarus, 260, p.320 – 331, doi:10.1016/j.icarus.2015.07.032.

Yu, Y. et al. (2018), The Dynamical Complexity of Surface Mass Shedding from a Top-shaped Asteroid Near the Critical Spin Limit, The Astronomical Journal, 156, 59, doi:10.3847/1538-3881/aacc7.

Zhang, Y. et al. (2017), Creep stability of the proposed AIDA mission target 65803 Didymos: I. Discrete cohesionless granular physics model, Icarus, 294, p.98 – 123, doi:10.1016/j.icarus.2017.04.027.

Zhang, Y. et al. (2018), Rotational Failure of Rubble-pile Bodies: Influences of Shear and Cohesive Strength, The Astrophysical Journal, 857:15, doi:10.3847/1538-4357/aab5b2.

Zimmerman, M. I. et al. (2016), Grain-scale supercharging and breakdown on airless regoliths, Journal of Geophysical Research: Planets, 121, p.2,150 – 2,165, doi:10.1002/2016JE005049.

**Figures and Tables**

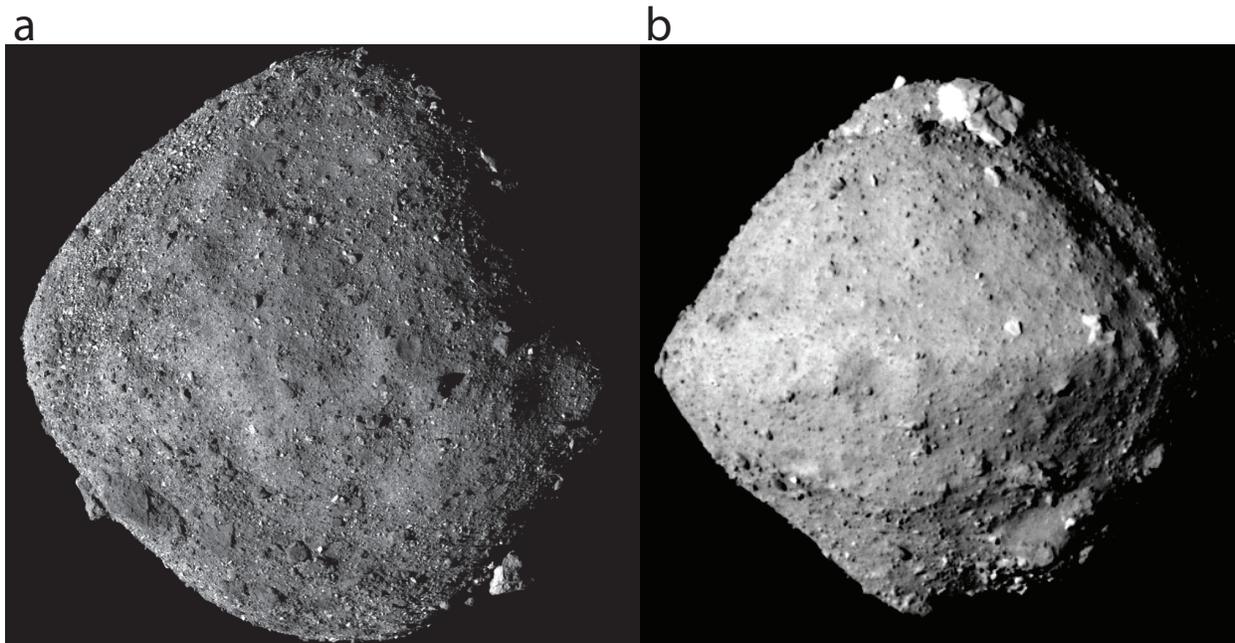

*Figure 1. Images of Bennu (Panel a) and Ryugu (Panel b). Panel a. Image credit: NASA/Goddard/University of Arizona. Panel b. Image credit: JAXA/UTokyo/Kochi U./Rikkyo U./Nagoya U./Chiba Ins. Tech/Meiji U./U. Aizu/AIST.*

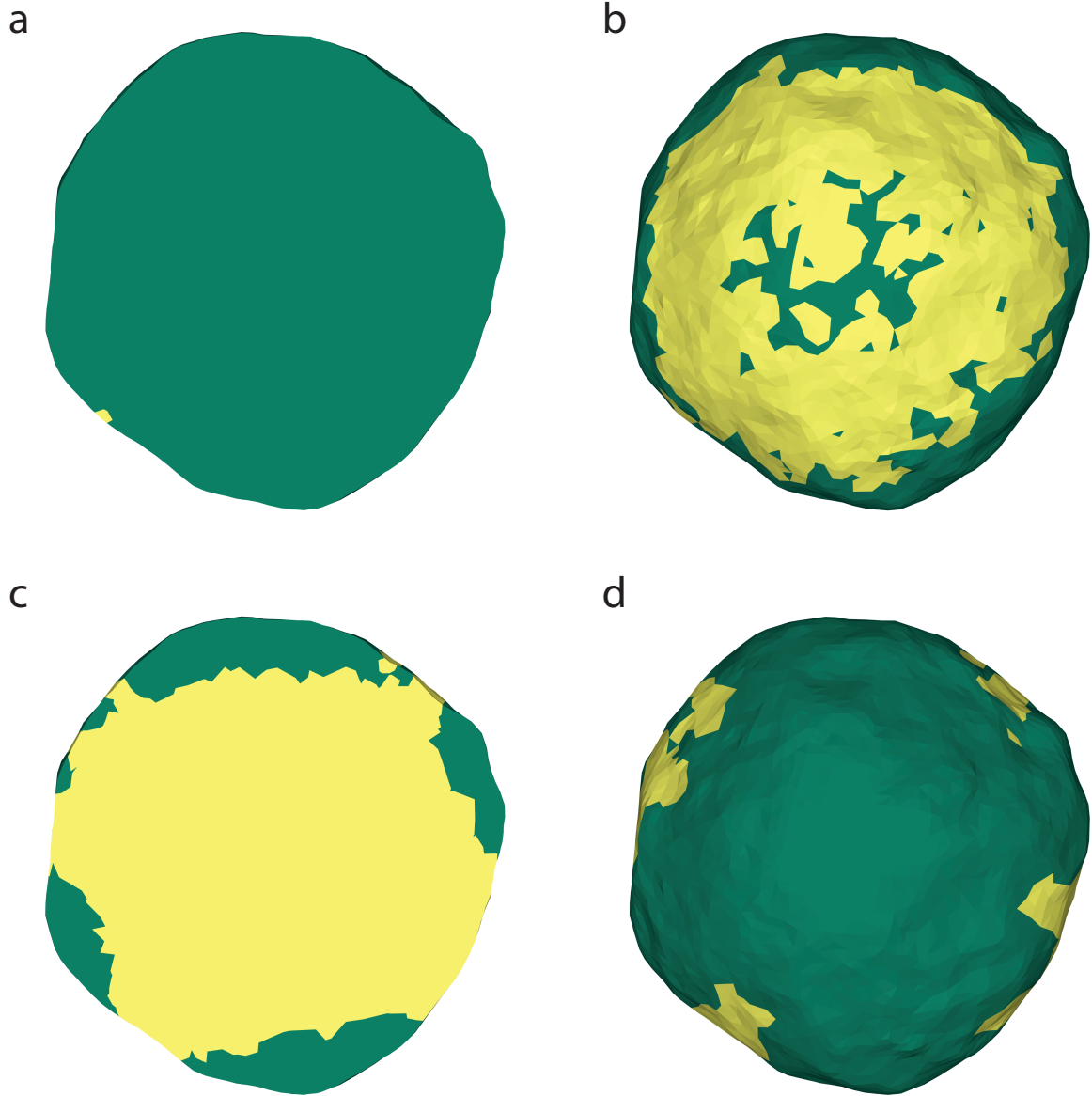

*Figure 2. Failed regions of Bennu at different spin periods. The areas in yellow describe structurally failed regions while those in green are intact, given a constant cohesive strength. Panels a and c are the sliced areas along the equatorial regions. Panels b and d show surface regions seen from the spin axes. Panels a and b show the case of the current spin period, 4.3 h, while Panels c and d give a spin period of 3.5 h [Scheeres et al., 2019]. The cohesive strength is fixed at 0.2 Pa for the 4.3 h case and at 0.8 Pa for the 3.5 h case.*

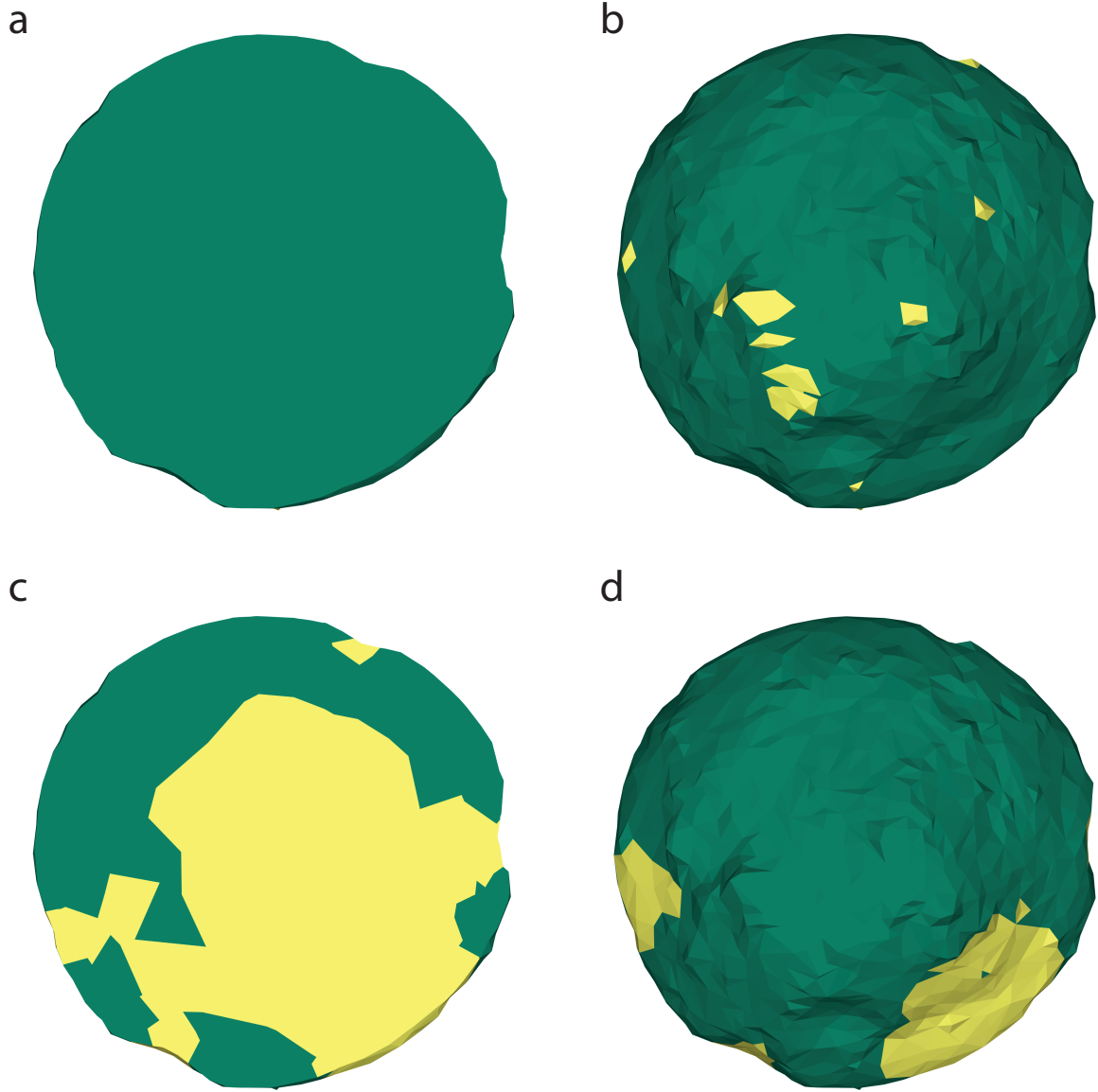

*Figure 3. Failed regions of Ryugu at different spin periods, viewed from the spin axis. The format of plot description is the same as in Figure 2. Panels a and b show the case of the current spin period, 7.6 h, while Panels c and d give a spin period of 3.5 h [Hirabayashi et al., 2019; Watanabe et al., 2019]. The cohesive strength is fixed at 1 Pa for the 7.6 h case and at 3.5 Pa for the 3.5 h case.*

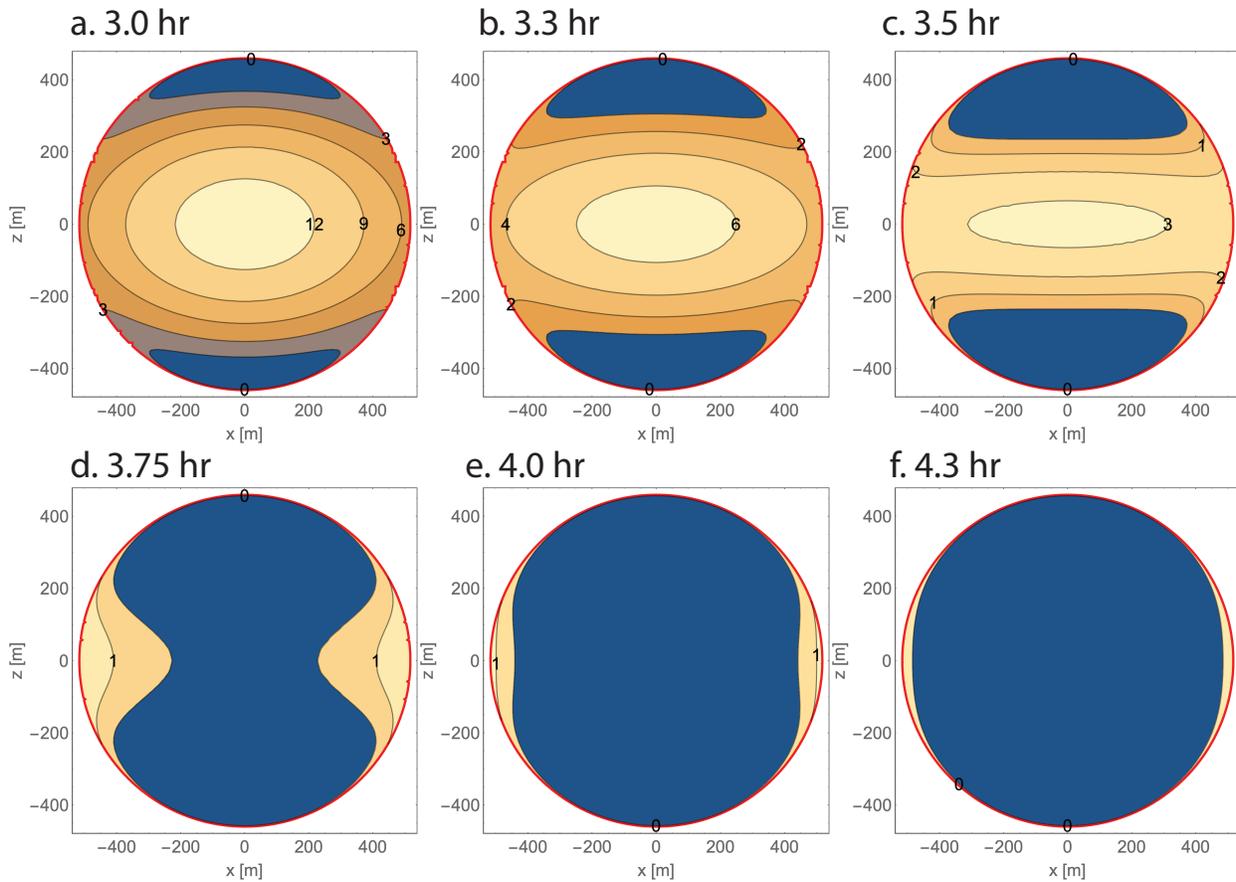

*Figure 4. Distribution of $Y_e^*$ on the slice along the spin axis of Bennu, derived by the semi-analytical model. The units of the contours are pascals.*

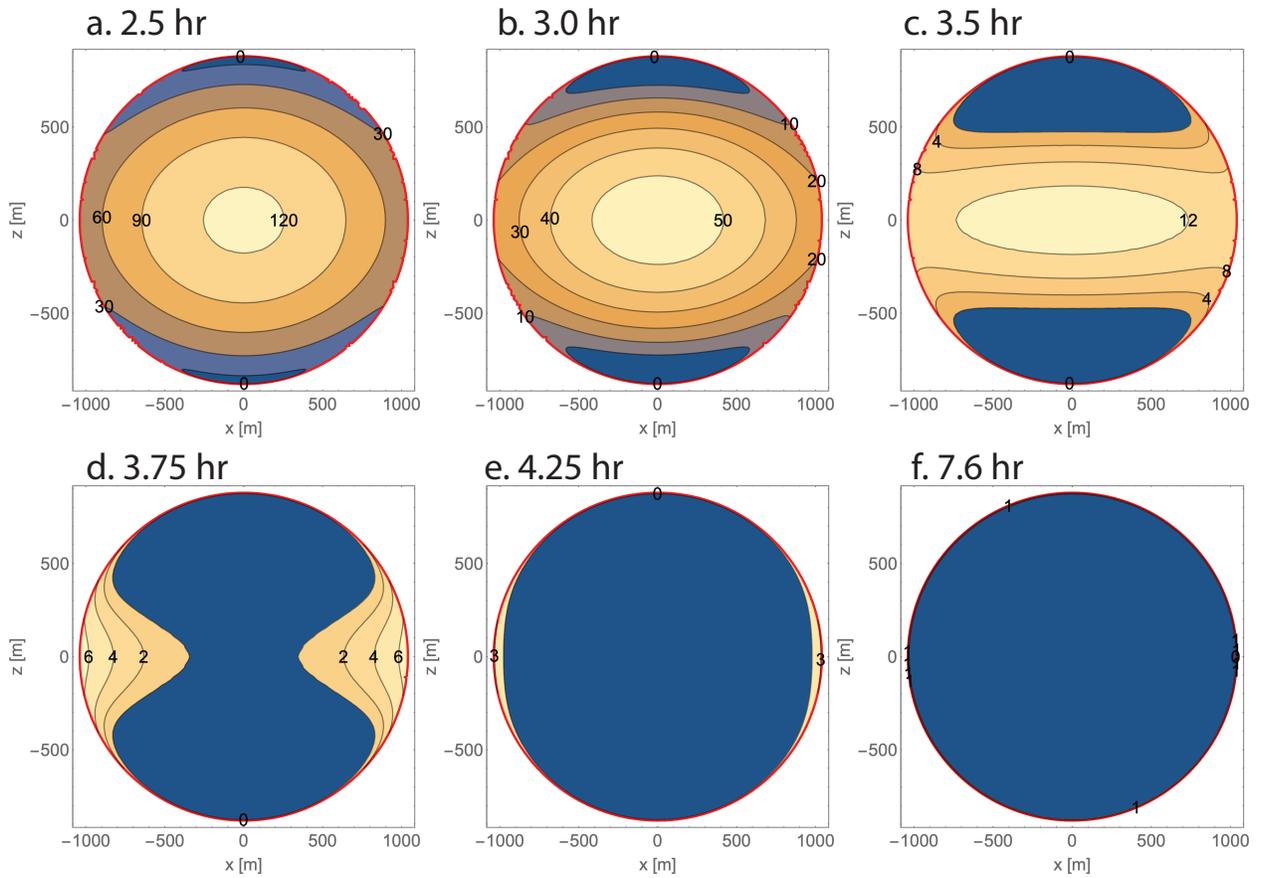

*Figure 5. Distribution of $Y_e^*$ on the slice along the spin axis of Ryugu, derived by the semi-analytical model. The units of the contours are pascals.*

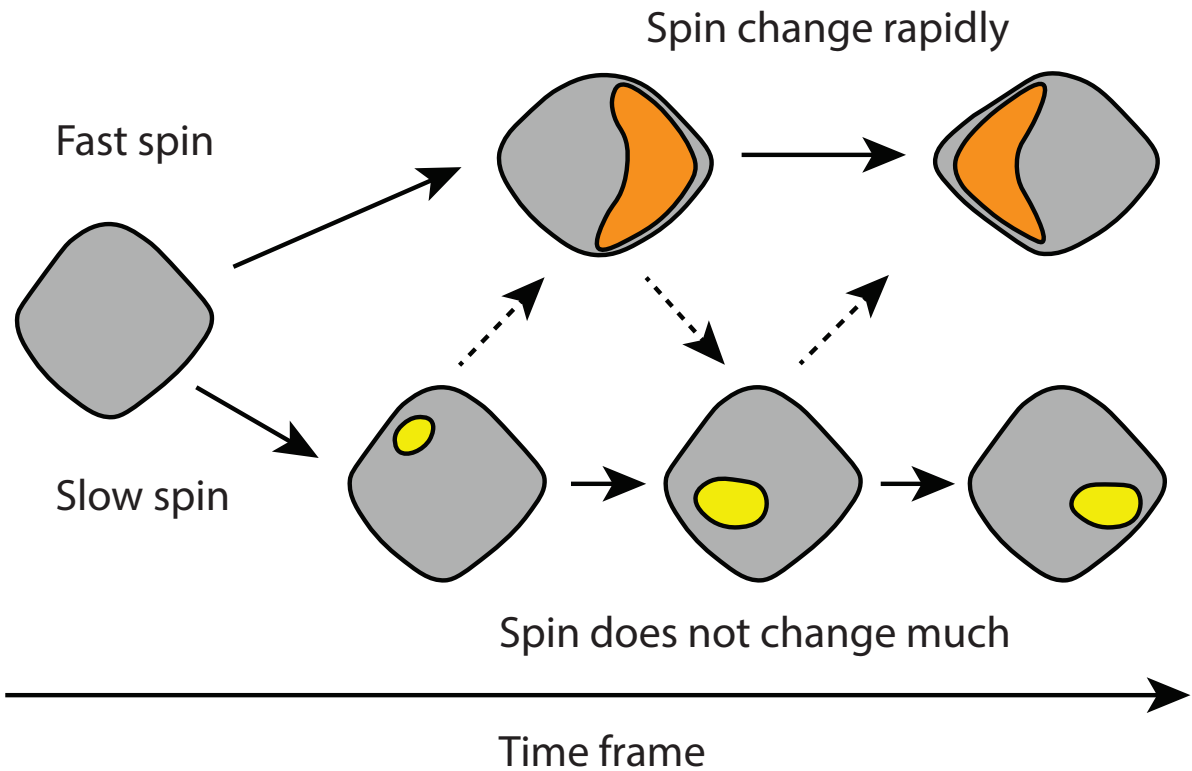

*Figure 6. Dependence of failure modes on spin states. The orange patches show large deformation caused by internal failure while the yellow ones describe small surface processes. With fast rotation, a top-shaped body may experience large deformation (the upper path). Slower rotation (but enough for surface failure) can induce small deformation (the lower path).*

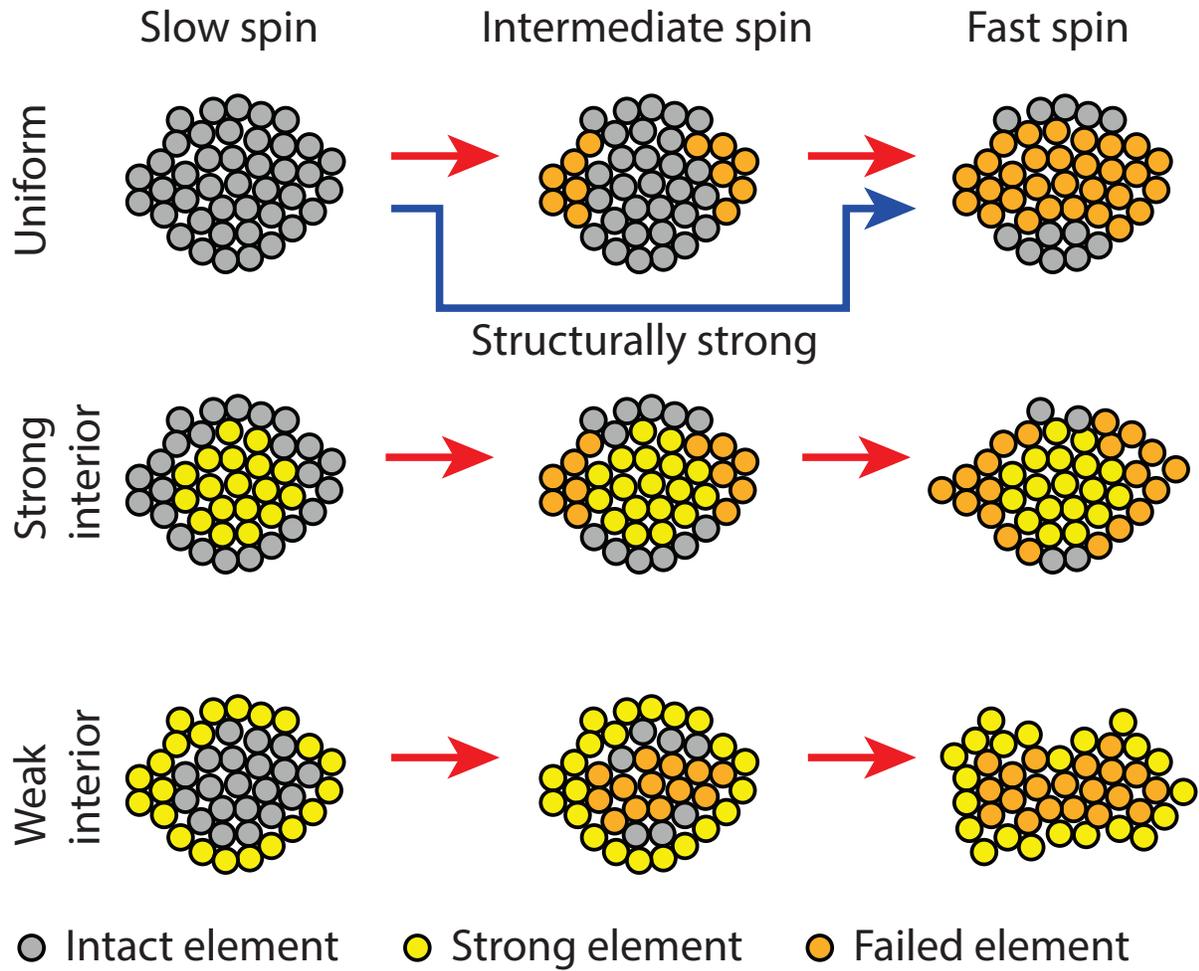

*Figure 7. Failure modes that depend on internal structure and spin. This figure is an extended version for the failure mode diagram of a spheroidal asteroid introduced by Hirabayashi [2015]. The case of a weak interior is added based on the study by Sánchez and Scheeres [2018].*